\documentclass[11pt]{article}
\usepackage{moriond,graphicx}

\def \bb{\overline{\cal B}}
\def \bea{\begin{eqnarray}}
\def \beq{\begin{equation}}

\def \eea{\end{eqnarray}}
\def \eeq{\end{equation}}
\def \gb{\overline{\Gamma}}
\def \gs{\stackrel{>}{\sim}}

\def \ob{\overline{B}^0}
\def \ok{\overline{K}^0}

\begin{document}
\vspace*{4cm}
\title{INTRODUCTION TO $B$ PHYSICS}

\author{ J. L. ROSNER }

\address{Enrico Fermi Institute and Department of Physics, University
of Chicago \\
5640 S. Ellis Avenue, Chicago, IL 60637 USA}

\maketitle\abstracts{
A short introduction to some current topics in $B$ physics is presented in
order to set the stage for some results to be announced at this Workshop.
After briefly reviewing the Cabibbo-Kobayashi-Maskawa (CKM) matrix and
information on its parameters, the decays of neutral $B$ mesons to several
CP eigenstates, and some $B \to \pi K$ decays, are discussed.  It is shown
that progress is being made on determination of all the angles of the
unitarity triangle.}

\section{Introduction}

The completion of a pioneering program on the physics of $B$ mesons at the
Cornell Electron Synchrotron (CESR) and the highly successful commissioning
of asymmetric $e^+ e^-$ colliders at PEP-II and KEK-B has led to a wealth of
data on the decays of $B$ mesons which can shed light on the weak and strong
interactions and on the violation of CP symmetry.  Hadron colliders are also
beginning to utilize specialized triggers to study $B$ mesons under conditions
of higher background but with the benefit of larger production cross sections.
One session of the present workshop is devoted to these results.  The present
Introduction seeks to put these exciting results in a broader context, showing
both what has been learned so far and what the future holds.

After reviewing the charge-changing weak transitions of quarks, described by
the Cabibbo-Kobayashi-Maskawa (CKM) matrix, in Sec.\ 2, we discuss several
decays of neutral $B$ mesons to CP eigenstates in Sec.\ 3, and a sampling
of $B \to \pi K$ decays in Sec.\ 4.  Sec.\ 5 concludes.

\section{The Cabibbo-Kobayashi-Maskawa Matrix}

The weak transitions between quarks with $Q = 2/3$ ($u,c,t$) and $Q = -1/3$
($d,s,b$) are encoded in a $3 \times 3$ unitary matrix $V$ known as the
Cabibbo-Kobayashi-Maskawa (CKM) matrix.\cite{Cab,KM}  The $tb$, $cs$, and
$ud$ elements of $V$ are of approximately unit magnitude, while $|V_{us}|
\simeq |V_{cd}| \simeq 0.22$, $|V_{cb}| \simeq |V_{ts}| \simeq 0.04$,
$|V_{td}| \simeq 0.008$, $|V_{ub}| \simeq 0.004$.  Phases in the last
two matrix elements can account for the observed CP violation in the kaon
system, as first noted by Kobayashi and Maskawa.\cite{KM}  One now seeks
tests of this mechanism through $B$ decays.  A convenient parametrization
of the CKM matrix, mentioned elsewhere in these Proceedings, is due to
Wolfenstein.\cite{WP}

\subsection{Unitarity}

\begin{figure}[t]
\begin{center}
\includegraphics[height=1.5in]{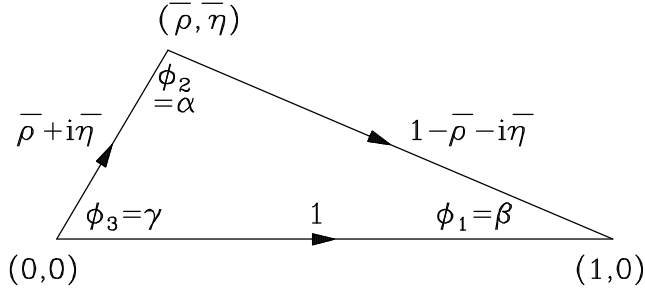}
\caption{The unitarity triangle.
\label{fig:ut}}
\end{center}
\end{figure}

The unitarity of the CKM matrix implies that the scalar product of any column
with the complex conjugate of any other column is zero, for example,
$V^*_{ub}V_{ud} + V^*_{cb} V_{cd} + V^*_{tb} V_{td} = 0$.  If one divides by
$-V^*_{cb} V_{cd}$, this relation becomes equivalent to a triangle in the
complex $\bar \rho + i \bar \eta$ plane, with vertices at (0,0) (angle $\phi_3
= \gamma$), (1,0) (angle $\phi_1 = \beta$), and $(\bar \rho, \bar \eta)$ (angle
$\phi_2 = \alpha$).  The triangle has unit base and its other two sides are
$\bar \rho + i \bar \eta = -(V^*_{ub}V_{ud}/ V^*_{cb} V_{cd})$ (opposite
$\phi_1 = \beta$) and $1 - \bar \rho - i \bar \eta = -(V^*_{tb}V_{td}/V^*_{cb}
V_{cd})$ (opposite $\phi_3 = \gamma$).  The result is shown in Fig.\
\ref{fig:ut}.

\subsection{Parameters}

Direct constraints on the CKM parameters are obtained from strange particle
decays ($V_{us} \simeq 0.22$), $b \to c$ decays ($V_{cb} \simeq 0.041$),
and $b \to u$ decays ($|V_{ub}/V_{cb}| \simeq 0.08 \sim 0.10$).  Indirect
constraints arise from various processes involving flavor-changing box
diagrams.  The $\bar s d \to \bar d s$ transition gives rise to CP-violating
$K^0$--$\ok$ mixing, whose magnitude (expressed through the parameter
$\epsilon$) imposes a constraint on Im($V_{td}^2$).  The $\bar b d \to \bar d
b$ transition generates $B^0$--$\ob$ mixing, from which a constraint on
$|V_{td}| \sim |1 - \bar \rho - i \bar \eta|$ may be obtained.  The $\bar b s
\to \bar s b$ transition generates $B_s$--$\overline{B}_s$ mixing; by comparing
the lower limit on this mixing with the observed $B^0$--$\ob$ mixing and
using estimates of flavor-SU(3)-symmetry breaking in matrix elements, one
finds $|V_{ts}/V_{td}| > 4.4$.  The combined impact of these constraints can
be conservatively summarized as requiring $0.08 \le \bar \rho \le 0.34$, $0.25
\le \bar \eta \le 0.43$,\cite{CKMf} though more restrictive analyses
(see, e.g., \cite{Ciu}) appear.

\section{$B^0$ Decays to CP Eigenstates}

Consider the decays $B^0 \to f$ (amplitude $A$) and $\ob \to f$
(amplitude $\bar A$), where $f$ is a CP eigenstate with eigenvalue $\xi_f = \pm
1$.  [See reviews \cite{BaBarPhys,TASI} for more details.]
As a result of $B^0$--$\ob$ mixing, a state which
is $B^0$ at proper time $t=0$ will evolve into one, denoted $B^0(t)$, which is
a mixture of $B^0$ and $\ob$.  Consequently, there will be two pathways to
the final state $f$:  one from $B^0$ through the amplitude $A$ and the other
from $\ob$ through the amplitude $\bar A$, which acquires an additional
phase $2 \phi_1$ through $B^0$--$\ob$ mixing.  The interference of these
two amplitudes can be different in the decay $B^0(t) \to f$ from that in
$\ob (t) \to f$, leading to a time-integrated rate asymmetry
\beq
{\cal A}_{CP} \equiv \frac{\Gamma(\ob \to f) - \Gamma(B^0 \to f)}
                          {\Gamma(\ob \to f) + \Gamma(B^0 \to f)}
\eeq
as well as to time-dependent rates
\beq
\left\{ \begin{array}{c} \Gamma[B^0(t) \to f] \\ \Gamma[\ob (t) \to f]
\end{array} \right\} \sim e^{- \Gamma t} [ 1 \mp {\cal A}_f \cos \Delta m t
 \mp {\cal S}_f \sin \Delta m t ]~~~,
\eeq
where
\beq
{\cal A}_f \equiv \frac{|\lambda|^2 - 1}{|\lambda|^2 + 1}~~,~~~
{\cal S}_f \equiv \frac{2 {\rm Im} \lambda}{|\lambda|^2 + 1}~~,~~~
\lambda \equiv e^{-2 i \phi_1} \frac{\bar A}{A}~~~.
\eeq
Note that one must have ${\cal S}_f^2 + {\cal A}_f^2 \le 1$.  I now discuss
specific cases.

\subsection{$B^0 \to J/\psi K_S$ and $\phi_1 = \beta$}

For this decay one has $\bar A/A \simeq \xi_{J/\psi K_S} = -1$.  One finds that
the time-integrated
asymmetry ${\cal A}_{CP}$ is proportional to $\sin(2 \phi_1)$.  Using this and
related decays involving the same $\bar b \to \bar s c \bar c$ subprocess,
BaBar \cite{Babeta} finds $\sin(2 \phi_1) = 0.741 \pm 0.067 \pm 0.033$
while Belle \cite{Bebeta} finds $\sin(2 \phi_1) = 0.719 \pm 0.074 \pm 0.035$.
The two values are quite consistent with one another; the world average
\cite{avbeta} is $\sin(2 \phi_1) = 0.734 \pm 0.054$, consistent with other
determinations.\cite{CKMf,Ciu,AL}

\subsection{$B^0 \to \pi^+ \pi^-$ and $\phi_2 = \alpha$}

Here the situation is more complicated because there are two competing
amplitudes contributing to the decay:  a ``tree'' $T$ and a ``penguin''
$P$.  The decay amplitudes are then
\beq
A = - (|T|e^{i \phi_3} + |P| e^{i \delta})~~,~~~
\bar A = - (|T|e^{-i \phi_3} + |P| e^{i \delta})~~~.
\eeq
The parameter $\delta$ is the relative $P/T$ strong phase.  The asymmetry
${\cal A}_{CP}$ would be proportional to $\sin(2 \phi_2)$ if the penguin
amplitude could be neglected.  However, it cannot, so methods have been
developed to deal with its contribution.

An isospin analysis \cite{GL} makes use of $B$ decays to $\pi^+ \pi^-$,
$\pi^\pm \pi^0$, and $\pi^0 \pi^0$ to separate the contributions of decays
involving $I=0$ and $I=2$ final states.  Information can then be obtained on
both strong and weak phases.  A potential problem with this method is that
the branching ratio of $B^0$ to $\pi^0 \pi^0$ may be very small, of order
$10^{-6}$.  I shall discuss instead methods \cite{GR02,GRconv} in which flavor
symmetry is used to estimate the magnitude of the penguin
amplitude.\cite{SW,GHLR,Charles}

The tree amplitude for $B^0 (= \bar b d) \to \pi^+ \pi^-$ involves the
subprocess $\bar b \to \pi^+ \bar u$, with the spectator $d$ quark combining
with $\bar u$ to form a $\pi^-$.  Its magnitude is $|T|$; its weak phase
is Arg($V^*_{ub}) = \phi_3$; by convention its strong phase is 0.  The
penguin amplitude involves the flavor structure $\bar b \to \bar d$, with the
final $\bar d d$ pair fragmenting into $\pi^+ \pi^-$.  Its magnitude is
$|P|$.  The dominant $t$ contribution in the loop diagram for $\bar b \to \bar
d$ can be integrated out and the unitarity relation $V_{td} V^*_{tb} =
- V_{cd} V^*_{cb} - V_{ud} V^*_{ub}$ used.  The $V_{ud} V^*_{ub}$ contribution
can be absorbed into a redefinition of the tree amplitude, after which
the weak phase of the penguin amplitude is 0 (mod $\pi$).  By definition, its
strong phase is $\delta$.

Knowledge of the time-dependent
asymmetries ${\cal S}_{\pi \pi}$ and ${\cal A}_{\pi \pi}$ suffices 
to specify both $\phi_3$ (or $\phi_2 = \pi - \phi_1 - \phi_3$) and $\delta$,
if one has an independent estimate of $|P/T|$.  This may be obtained by using
flavor SU(3) to get $|P|$ from $B^+ \to K^0 \pi^+$ \cite{SW,GHLR,GR95} and
factorization to get $|T|$ from $B \to \pi l \nu$.\cite{LR}  Since the
$T$ amplitude obtained from factorization is not precisely the same as that
which contains a small $V_{ud} V^*_{ub}$ contribution from the penguin,
an alternative method \cite{GRconv,Charles} makes direct use of the measured
ratio of the $B^+ \to K^0 \pi^+$ and $B^0 \to \pi^+ \pi^-$ branching ratios
to constrain $|P/T|$.  I shall discuss the first method since it is simpler.

\begin{figure}[t]
\begin{center}
\includegraphics[height=4.5in]{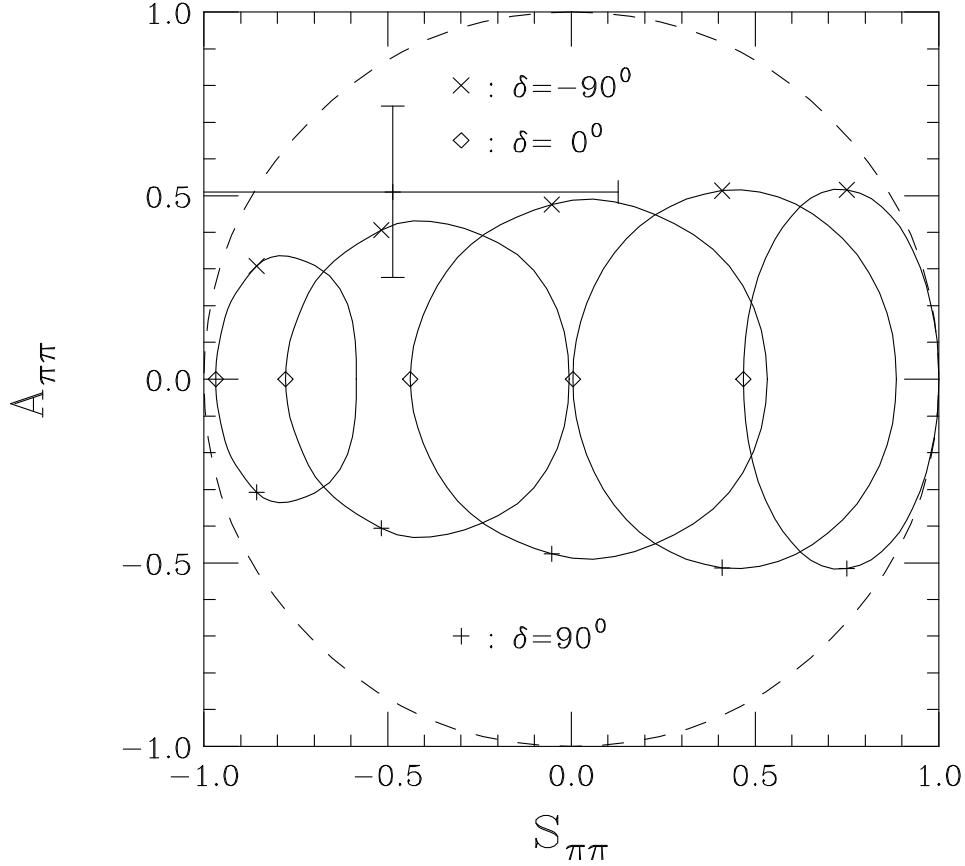}
\caption{Curves depicting dependence of ${\cal S}_{\pi \pi}$ and $
{\cal A}_{\pi \pi}$ on $\delta$ ($- \pi \le \delta \le \pi$).
From right to left the curves correspond to $\phi_2 = (120^\circ, 105^\circ,
90^\circ, 75^\circ, 60^\circ)$.  Plotted point:  average of BaBar
and Belle values (see text). 
\label{fig:sa}}
\end{center}
\end{figure}

In addition to ${\cal S}_{\pi \pi}$ and ${\cal A}_{\pi \pi}$, a useful quantity
is the ratio of the $B^0 \to \pi^+ \pi^-$ branching ratio $\bb(\pi^+ \pi^-)$
(averaged over $B^0$ and $\ob$) to that due to the tree amplitude alone:
\beq
R_{\pi \pi} \equiv \frac{\bb(\pi^+ \pi^-)}{\bb(\pi^+ \pi^-)|_{\rm tree}}
= 1 - 2 \left| \frac{P}{T} \right| \cos \delta \cos(\phi_1 + \phi_2)
+ \left| \frac{P}{T} \right|^2~~~.
\eeq
One also has
\beq
R_{\pi \pi} {\cal S}_{\pi \pi} = \sin 2 \phi_2 + 2 \left| \frac{P}{T} \right|
\cos \delta \sin(\phi_1 - \phi_2) - \left| \frac{P}{T} \right|^2 \sin(2 \phi_1)
~~~,
\eeq
\beq
R_{\pi \pi} {\cal A}_{\pi \pi} = - 2 |P/T| \sin \delta \sin (\phi_1 + \phi_2)
~~~.
\eeq
The value of $\phi_1 = \beta$ is specified to within a few degrees; we shall
take it to have its central value $\phi_1 = 23.6^\circ$.  The value of
$|P/T|$ (updating \cite{GR02,GRconv}) is $0.28 \pm 0.06$.  Taking the central
value, we plot trajectories in the (${\cal S}_{\pi \pi},{\cal A}_{\pi \pi}$)
plane as $\delta$ is allowed to vary from $- \pi$ to $\pi$.  The result is
shown in Fig.\ \ref{fig:sa}.

The experimental situation regarding the time-dependent asymmetries is not
yet settled.  As shown in Table \ref{tab:sa}, BaBar \cite{Bapipi} and Belle
\cite{Bepipi} obtain very different values, especially for ${\cal S}_{\pi
\pi}$.  Even if this conflict were to be resolved, however, one sees the
possibility of a discrete ambiguity, since curves for different values of
$\phi_2$ intersect one another.

\begin{table}
\caption{Values of ${\cal S}_{\pi \pi}$ and ${\cal A}_{\pi \pi}$ quoted by
BaBar and Belle and their averages.  Here we have applied scale factors
$S \equiv \sqrt{\chi^2} = (2.31,1.24)$ to the errors for
${\cal S}_{\pi \pi}$ and ${\cal A}_{\pi \pi}$, respectively.
\label{tab:sa}}
\begin{center}
\begin{tabular}{|c|c|c|c|} \hline
    Quantity         & BaBar \cite{Bapipi}  & Belle \cite{Bepipi}            &
    Average \\ \hline
${\cal S}_{\pi \pi}$ & $0.02\pm0.34\pm0.05$ & $-1.23\pm0.41^{+0.08}_{-0.07}$ &
    $-0.49 \pm 0.61$ \\ 
${\cal A}_{\pi \pi}$ & $0.30\pm0.25\pm0.04$ & $ 0.77 \pm 0.27 \pm 0.08$ &
    $0.51 \pm 0.23$ \\ \hline
\end{tabular}
\end{center}
\end{table}

\begin{figure}[t]
\begin{center}
\includegraphics[height=3in]{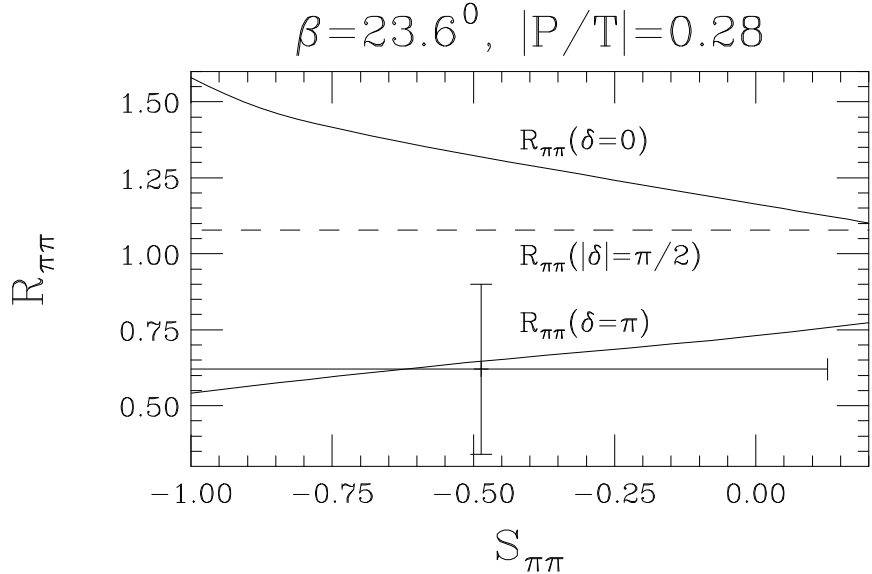}
\caption{Curves depicting dependence of $R_{\pi \pi}$ on ${\cal S}_{\pi \pi}$
for various values of $\delta$.  The plotted point is the average of BaBar
and Belle values for $ {\cal S}_{\pi \pi}$ (see text). 
\label{fig:sr}}
\end{center}
\end{figure}

The discrete ambiguity may be resolved with the help of $R_{\pi \pi}$.  The
most recent average of branching ratios, including ones from Belle presented
at this Conference,\cite{Bebrs} yields $\bb(B^0 \to \pi^+ \pi^-) =
(4.55 \pm 0.44) \times 10^{-6}$ (see below), while \cite{LR} $\bb(B^0 \to \pi^+
\pi^-)|_{\rm tree} = (7.3 \pm 3.2) \times 10^{-6}$, so that $R_{\pi \pi} = 0.62
\pm 0.28$.  One can then plot $R_{\pi \pi}$ as a function of ${\cal S}_{\pi
\pi}$ for any value of $\delta$; some examples are shown in Fig. \ref{fig:sr}.

If the errors on $R_{\pi \pi}$ can be reduced to $\pm 0.1$, a distinction
between $\delta = 0$ and $\delta = \pm \pi$ will be possible.
One possibility for improving this situation
would be to measure $(d \sigma/ d q^2)(B \to \pi l \nu)$ at $q^2 = m_\pi^2$
and to obtain $|T|$ using factorization.\cite{LR}  A value of $|T|$ obtained
\cite{xiao} from $B^+ \to \pi^+ \pi^0$ implies $\bb(B^0 \to \pi^+ \pi^-)|_{\rm
tree} = (9.0 \pm 1.8) \times 10^{-6}$ and thus would favor $R_{\pi \pi}$
significantly below 1.  However, this work underestimates the uncertainty
due to the color-suppressed contribution to $B^+ \to \pi^+ \pi^0$.\cite{MN}
A value of $R_{\pi \pi}$ below 1 would favor large $\delta$ (unexpected in
factorization approaches) and, referring to Fig.\ \ref{fig:sa}, larger values
of $\phi_2$.  Such a conclusion, in my opinion, is premature.

\subsection{$B^0 \to \phi K_S$:  New Physics?}

In the decay $B^0 \to \phi K_S$, governed by the $\bar b \to \bar s$ penguin
amplitude, the standard model predicts the same CP asymmetries as in those
processes (like the ``golden'' $J/\psi K_S$ mode) governed by the $\bar b \to
\bar s c \bar c$ tree amplitude.  In both cases the weak phase is expected to
be 0 (mod $\pi$), so the indirect CP asymmetry should be governed entirely
by $B^0$--$\ob$ mixing and thus should be proportional to $\sin 2 \phi_1$.
There should be no direct CP asymmetries (i.e., one expects ${\cal A} \simeq
0$) in either case.  This is certainly true for $B \to J/\psi K$; ${\cal A}$
is consistent with zero in the neutral mode, while the direct CP asymmetry is
consistent with zero in the charged mode.\cite{Babeta}  However, a different
result for $B^0 \to \phi K_S$ could point to new physics in the $\bar b \to
\bar s$ penguin amplitude.\cite{GW}

The experimental situation for the asymmetries in $B^0 \to \phi K_S$ is
shown in Table \ref{tab:phks}.  We have included an updated result presented
by the BaBar Collaboration at this conference.\cite{Baphks}  One cannot
conclude much about ${\cal A}_{\phi K_S}$ as a result of the substantial
discrepancy between BaBar and Belle.  However, the value of ${\cal S}_{\phi
K_S}$, which should be equal to $\sin 2 \phi_1 = 0.734 \pm 0.054$ in the
standard model, is about $2.7 \sigma$ away from it.  If one assumes that
the amplitudes for $B^0 \to \phi K^0$ and $B^+ \to \phi K^+$ are equal, as
is true in many approaches, one expects the time-integrated CP asymmetry
$A_{CP}$ in the charged mode to be equal to ${\cal A}_{\phi K_S}$.  The
BaBar Collaboration \cite{Aubert:2003tk} has recently reported
$A_{CP} = 0.039 \pm 0.086 \pm 0.011$.

\begin{table}
\caption{Values of ${\cal S}_{\phi K_S}$ and ${\cal A}_{\phi K_S}$ quoted by
BaBar and Belle and their averages.  Here we have applied a scale factor of
$\sqrt{\chi^2} = 2.29$ to the error on ${\cal A}_{\phi K_S}$.
\label{tab:phks}}
\begin{center}
\begin{tabular}{|c|c|c|c|} \hline
    Quantity         & BaBar \cite{Baphks}  & Belle \cite{Bephks}            &
    Average \\ \hline
${\cal S}_{\phi K_S}$ & $-0.18\pm0.51\pm0.07$ & $-0.73\pm0.64\pm0.22$ &
 $-0.38 \pm 0.41$ \\ 
${\cal A}_{\phi K_S}$ & $0.80\pm0.38\pm0.12$ & $-0.56\pm0.41\pm0.16$ &
 $0.19 \pm 0.68$ \\ \hline
\end{tabular}
\end{center}
\end{table}

Many proposals for new physics can account for such a discrepancy.\cite{npphks}
I describe a method similar to that \cite{GR02,GRconv} used in
analyzing $B^0 \to \pi \pi$ for extracting a new physics amplitude, developed
in collaboration with Cheng-Wei Chiang.\cite{CR03}  One uses the measured
values of ${\cal S}_{\phi K_S}$ and ${\cal A}_{\phi K_S}$ and the ratio
\beq \label{eqn:rphks}
R_{\phi K_S} \equiv \frac{\bb(B^0 \to \phi K_S)}{\bb(B^0 \to \phi K_S)|_{\rm
 std}} = 1 + 2 r \cos \phi \cos \delta + r^2~~~,
\eeq
where $r$ is the ratio of the magnitude of the new amplitude to the one in
the standard model, and $\phi$ and $\delta$ are their relative weak and
strong phases.  For any values of $R_{\phi K_S}$, $\phi$, and $\delta$, Eq.\
(\ref{eqn:rphks}) can be solved for the amplitude ratio $r$ and one then
calculates
\bea
R_{\phi K_S} {\cal S}_{\phi K_S} & = & \sin 2 \phi_1 + 2 r \cos \delta
\sin(2 \phi_1 - \phi) + r^2 \sin 2(\phi_1 - \phi)~~\\
R_{\phi K_S} {\cal A}_{\phi K_S} & = & 2 r \sin \phi \sin \delta~~~.
\eea
The $\phi K_S$ branching ratio in the standard model is calculated using the
penguin amplitude from $B^+ \to K^{*0} \pi^+$ and an estimate of electroweak
penguin corrections.  It was found \cite{CR03} that $R_{\phi K_S} = 1.0 \pm
0.2$.

\begin{figure}
\begin{center}
\includegraphics[height=4.5in]{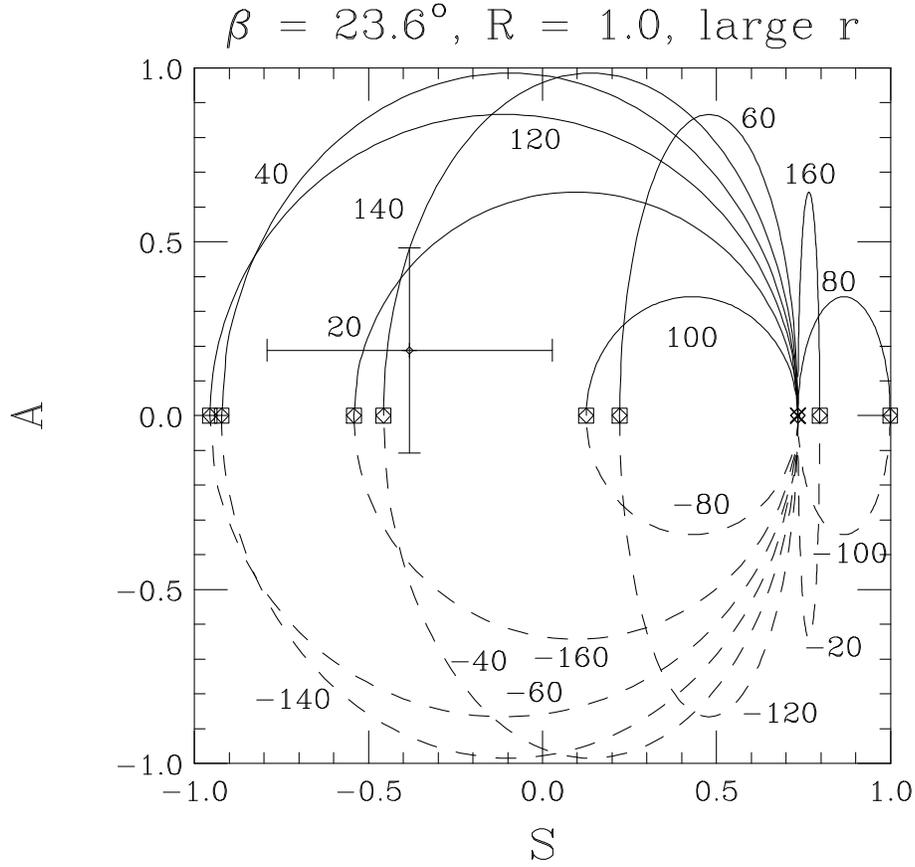}
\caption{Curves depicting dependence of ${\cal S}_{\phi K_S}$ and ${\cal A}_
{\phi K_S}$ on $\delta$ ($0 \le \delta \le \pi$).  The plot is for $\beta =
23.6^{\circ}$ and $R=1$ (choosing only the non-zero solution for $r$).
Curves are labeled by values of $\phi$ (dashed: $\phi < 0$; solid: $\phi >
0$) in degrees.  Squares and diamonds correspond to values of $\delta = 0$ or
$\pi$.  The point at ${\cal S}_{\phi K_S} = 0.734$, ${\cal A}_{\phi K_S} = 0$
corresponds to $\phi = 0,~\pm \pi$ for all $\delta$.  The plotted data point
is based on the the averages quoted in Table \ref{tab:phks} but without the
scale factor for ${\cal A}_{\phi K_S}$.
\label{fig:saphks}}
\end{center}
\end{figure}

For values of $\phi$ between $-\pi$ and $\pi$, curves of ${\cal S}_{\phi K_S}$
vs.\ ${\cal A}_{\phi K_S}$ are plotted in Fig.\ \ref{fig:saphks} as $\delta$
varies from $0$ to $\pi$.  Both ${\cal S}$ and ${\cal A}$ are unchanged under
$\phi \to \phi + \pi$, $\delta \to \delta - \pi$, while ${\cal S} \to {\cal
S}$, ${\cal A} \to - {\cal A}$ under $\phi \to \phi + \pi$, $\delta \to \pi -
\delta$.

\begin{figure}
\begin{center}
\includegraphics[height=2.9in]{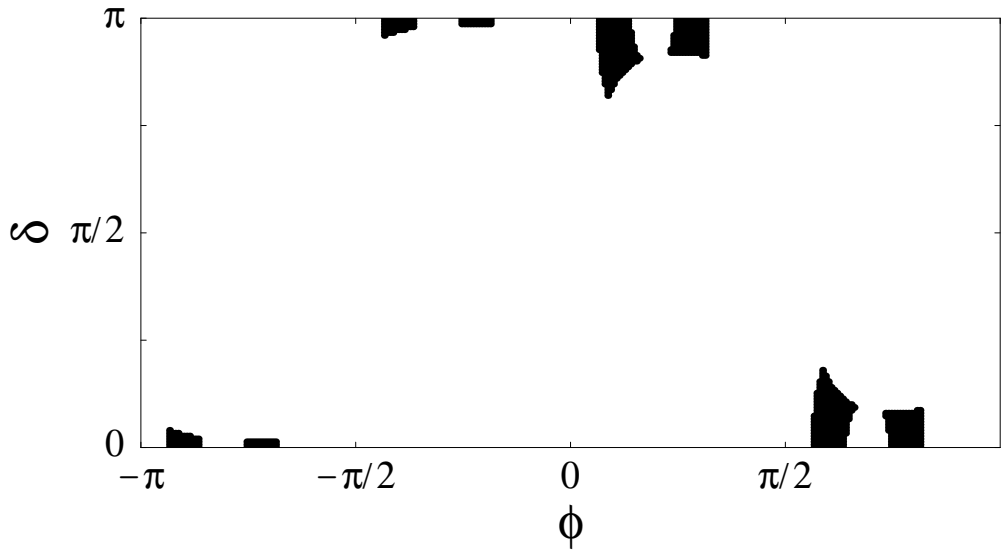} \\
\caption{Allowed regions in the $\phi$-$\delta$ plane for $R = 1$.  Note the
symmetry under $\delta \to \pi - \delta$ and $\phi \to \phi + \frac {\pi}{2}$.
\label{fig:phidel}}
\end{center}
\end{figure}

Various regions of $(\phi, \delta)$ can reproduce the observed values of
${\cal S}_{\phi K_S}$ and ${\cal A}_{\phi K_S}$.  Some of these are shown in
Fig.\ \ref{fig:phidel}, while others correspond to shifts in $\phi$
and $\delta$ by $\pm \pi$.  As errors on the observables shrink, so will the
allowed regions.  However, as Fig.\ \ref{fig:saphks} makes clear, there will
always be a solution for {\it some} $\phi$ and $\delta$ as long as $R$
remains compatible with 1.  (The allowed regions of $\phi$ and $\delta$ are
restricted if $R \ne 1$.\cite{CR03})  Typical values of $r$ are of order 1;
one generally needs to invoke new-physics amplitudes comparable to those in
the standard model.

The above scenario envisions new physics entirely in $B^0 \to \phi K^0$ and
not in $B^+ \to K^{*0} \pi^+$.  An alternative is that new physics
contributes to the $\bar b \to \bar s$ penguin amplitude and thus appears
in {\it both} decays.  Here it is convenient to define a ratio
\beq
R' \equiv \frac{\gb(B^0 \to \phi K^0)}{\gb(B^+ \to K^{*0} \pi^+)}~~~,
\eeq
where $\gb$ denotes a partial width averaged over a process and its CP
conjugate.  Present data indicate $R' = 0.78 \pm 0.17$.  The $B^0 \to \phi K^0$
amplitude contains a contribution from both the gluonic and electroweak penguin 
terms, while $B^+ \to K^{*0} \pi^+$ contains only the former.  Any departure
from the expected ratio of the electroweak to gluonic penguin amplitudes
would signify new physics.  Again, the central value of ${\cal S}$ would
suggest this to be the case.\cite{CR03}

\subsection{$B^0 \to \eta' K_S$:  No New Physics Needed}

At present neither the rate nor the CP asymmetry in $B \to \eta' K$ present
a significant challenge to the standard model.  The rate can be reproduced
with the help of a modest contribution from a ``flavor-singlet penguin''
amplitude, the need for which was pointed out \cite{GR95,DGR95} prior to the
observation of this decay.  One only needs to boost the standard penguin
amplitude's contribution by about 50\% via the flavor-singlet term in order to
explain the observed rate.\cite{DGR97,CR01,FHH}.  (Ref.\ \cite{BN} instead
finds an enchanced standard-penguin contribution to $\eta'$ production.)
The CP asymmetry is
not a problem; the ordinary and singlet penguin amplitudes are expected
to have the same weak phase Arg$(V^*_{ts}V_{tb}) \simeq \pi$ and hence one
expects ${\cal S}_{\eta' K_S} \simeq \sin 2 \phi_1$, ${\cal A}_{\eta' K_S}
\simeq 0$.  The experimental situation is shown in Table \ref{tab:etapks}.
The value of ${\cal S}_{\eta' K_S}$ is consistent with the standard model
expectation at the $1 \sigma$ level, while ${\cal A}_{\eta' K_S}$ is consistent
with zero.

\begin{table}
\caption{Values of ${\cal S}_{\eta' K_S}$ and ${\cal A}_{\eta' K_S}$ quoted by
BaBar and Belle and their averages.  Here we have applied scale factors
$S \equiv \sqrt{\chi^2} = (1.48,1.15)$ to the errors for
${\cal S}_{\eta' K_S}$ and ${\cal A}_{\eta' K_S}$, respectively.
\label{tab:etapks}}
\begin{center}
\begin{tabular}{|c|c|c|c|} \hline
    Quantity         & BaBar \cite{Baphks}  & Belle \cite{Bephks}            &
    Average \\ \hline
${\cal S}_{\eta' K_S}$ & $0.02\pm0.34\pm0.03$ & $0.76\pm0.36^{+0.05}_{-0.06}$ &
$0.37 \pm 0.37$  \\ 
${\cal A}_{\eta' K_S}$ & $-0.10\pm0.22\pm0.03$ & $0.26\pm0.22\pm0.03$ &
$0.08 \pm 0.18$  \\ \hline
\end{tabular}
\end{center}
\end{table}

The singlet penguin amplitude may contribute elsewhere in $B$ decays.  It is
a possible source of a low-effective-mass $\bar p p$ enhancement
\cite{Kpp} in $B^+ \to \bar p p K^+$.\cite{JRbbbar}

\section{$B \to \pi K$ Decays and $\phi_3 = \gamma$}

The decays $B \to K \pi$ (with the exception of $B^0 \to K^0 \pi^0$) are
self-tagging.  For example, the $K^+ \pi^-$ final state is expected to
originate purely from a $B^0$ and not from a $\ob$.  Since such self-tagging
decays do not involve a CP eigenstate, one must contend with both weak and
strong phases.  Nonetheless several methods permit one to separate these
from one another.  We give two examples below.

\subsection{$B^0 \to K^+ \pi^-$ vs.\ $B^+ \to K^0 \pi^+$}

The decay $B^+ \to K^0 \pi^+$ is a pure penguin ($P$) process, while the
amplitude for $B^0 \to K^+ \pi^-$ is proportional to $P + T$, where $T$ is a
(strangeness-changing) tree amplitude.  The ratio $T/P$ has magnitude $r$, weak
phase $\phi_3 \pm \pi = \gamma \pm \pi$, and strong phase $\delta$.  The ratio
$R_0$ of these two rates (averaged over a process and its CP conjugate) is
\beq \label{eqn:Rval}
R_0 \equiv \frac{\gb(B^0 \to K^+ \pi^-)}{\gb(B^+ \to K^0 \pi^+)} =
1 - 2 r \cos \gamma \cos \delta + r^2 \ge \sin^2 \gamma~~~,
\eeq
where the inequality holds for any $r$ and $\delta$.  If $R_0$ were
significantly
less than $1$ this inequality could be used to impose a useful constraint on
$\gamma$.\cite{FM}  On the basis of the latest branching ratios from
BaBar,\cite{Babrs} Belle,\cite{Bebrs} and CLEO,\cite{CLbrs} summarized
in Table \ref{tab:brs}, using the $B^+/B^0$ lifetime ratio $\tau_+/\tau_0
= 1.073 \pm 0.014$,\cite{LEPBOSC}, one finds $R_0 = 0.99 \pm 0.09$, which is
consistent with 1 and does not permit application of the bound.  However,
using additional information on $r$ and the CP asymmetry in $B^0 \to K^+
\pi^-$, one can obtain a constraint on $\gamma$.\cite{GR02,GRKpi}

\begin{table}
\caption{Branching ratios for some charmless two-body $B$ decays, in units of
$10^{-6}$.
\label{tab:brs}}
\begin{center}
\begin{tabular}{|c|c|c|c|c|} \hline
Mode & BaBar \cite{Babrs} & Belle \cite{Bebrs} & CLEO \cite{CLbrs} & Average \\
\hline
$K^+ \pi^-$ & $17.9 \pm 0.9 \pm 0.6$ & $18.5 \pm 1.0 \pm 0.7$ &
 $18.0^{+2.3+1.2}_{-2.1-0.9}$ & $18.15 \pm 0.77$ \\
$K^+ \pi^0$ & $12.8 \pm 1.2 \pm 1.0$ & $12.8 \pm 1.4^{+1.4}_{-1.0}$ &
 $12.9^{+2.4+1.2}_{-2.2-1.1}$ & $12.82 \pm 1.09$ \\
$K^0 \pi^+$ & $17.5 \pm 1.8 \pm 1.3$ & $22.0 \pm 1.9 \pm 1.1$ &
 $18.8^{+3.7+2.1}_{-3.3-1.8}$ & $19.65 \pm 1.45$ \\
$K^0 \pi^0$ & $10.4 \pm 1.5 \pm 0.8$ & $12.6 \pm 2.4 \pm 1.4$ &
 $12.8^{+4.0+1.7}_{-3.3-1.4}$ & $11.21 \pm 1.36$ \\ \hline
$\pi^+ \pi^-$ & $4.7 \pm 0.6 \pm 0.2$ & $4.4 \pm 0.6 \pm 0.3$ &
 $4.5^{+1.4+0.5}_{-1.2-0.4}$ & $4.55 \pm 0.44$ \\
$\pi^+ \pi^0$ & $5.5 \pm 1.0 \pm 0.6$ & $5.3 \pm 1.3 \pm 0.5$ &
 $4.6^{+1.8+0.6}_{-1.6-0.7}$ & $5.26 \pm 0.80$ \\ \hline
\end{tabular}
\end{center}
\end{table}

\begin{figure}
\begin{center}
\includegraphics[height=3.5in]{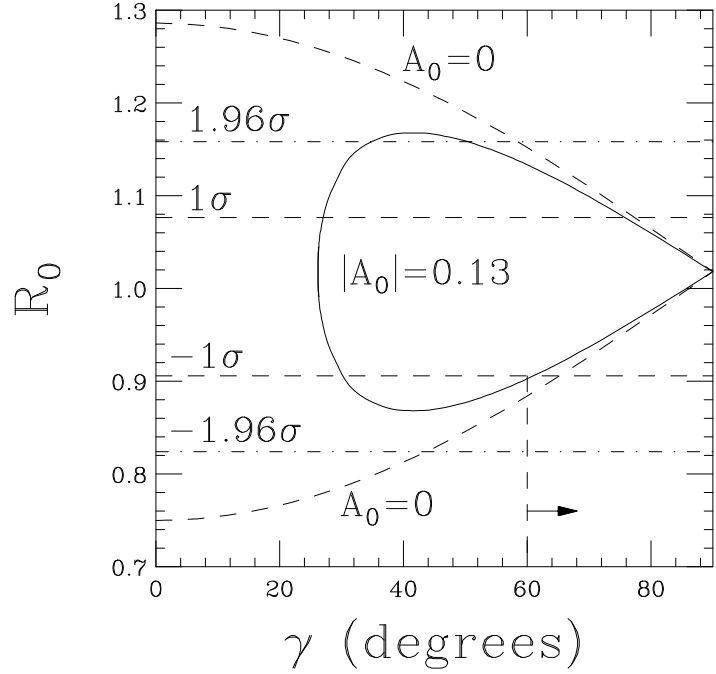}
\caption{Behavior of $R_0$ for $r = 0.134$ and $A_0 = 0$ (dashed curves) or
$|A_0| = 0.13$ (solid curve) as a function of the weak phase $\gamma$.
Horizontal dashed lines denote $\pm 1 \sigma$ experimental limits on $R$,
while dot-dashed lines denote $95\%$ c.l. ($\pm 1.96 \sigma$) limits.
The upper branches of the curves correspond to the case $\cos \gamma
\cos \delta <0$, while the lower branches correspond to $\cos \gamma
\cos \delta >0$.
\label{fig:R}}
\end{center}
\end{figure}

Define a ``pseudo-asymmetry'' normalized by the rate for $B^0 \to K^0
\pi^+$, a process which is expected not to display a CP asymmetry
since only the penguin amplitude contributes to it:
\beq \label{eqn:asy}
A_0 \equiv \frac{\Gamma(\ob \to K^- \pi^+) - \Gamma(B^0 \to K^+ \pi^-)}
{2 \gb(B^+ \to K^0 \pi^+)} = - 2 r \sin \gamma \sin \delta~~~.
\eeq
One may eliminate $\delta$ between this equation and Eq.\ (\ref{eqn:Rval})
and plot $R_0$ as a function of $\gamma$ for the allowed range of $|A_0|$.  One
needs an estimate of $r$, whose present value (based on the branching ratios
in Table \ref{tab:brs} and arguments given in Refs.\ \cite{GR02,GRKpi})
is $r = 0.17 \pm 0.04$.  Here one must take the value of $T$ from $B \to
\pi l \nu$,\cite{LR} using flavor SU(3) to relate the strangeness-preserving
and strangeness-changing terms.  The latest BaBar and Belle data imply
$A_0 = -0.088 \pm 0.040$, leading us to take $|A_0| \le 0.13$ at the $1 \sigma$
level.  Curves for $A_0=0$ and $|A_0| = 0.13$ are shown in Fig.\ \ref{fig:R}.
The lower limit $r = 0.134$ is used to generate these curves since the limit
on $\gamma$ will be the most conservative.

At the $1 \sigma$ level, using the constraints that $R_0$ must lie between 0.90
and 1.08 and $|A_0|$ must lie between zero and 0.13, one can establish the bound
$\gamma \gs 60^\circ$.  No bound can be obtained at the 95\% confidence level,
however.  Despite the impressive improvement in experimental precision (a
factor of 2 decrease in errors since the analysis of Ref.\ \cite{GR02}),
further data are needed in order for a useful constraint to be obtained.

\subsection{$B^+ \to K^+ \pi^0$ vs. $B^+ \to K^0 \pi^+$}

The comparison of rates for $B^+ \to K^+ \pi^0$ and $B^+ \to K^0 \pi^+$ also
can give information on $\phi_3 = \gamma$.  The amplitude for $B^+ \to K^+
\pi^0$ is proportional to $P + T + C$, where $C$ is a color-suppressed
amplitude.  Originally it was suggested that this amplitude be compared
with $P$ from $B^+ \to K^0 \pi^+$ and $T+C$ taken from $B^+ \to \pi^+ \pi^0$
using flavor SU(3),\cite{GRL} using a triangle construction to determine
$\gamma$.  However, electroweak penguin amplitudes contribute significantly
in the $T+C$ term.\cite{EWP}  It was noted subsequently \cite{NR}
that since the $T+C$ amplitude corresponds to isospin $I(K \pi) = 3/2$
for the final state, the strong-interaction phase of its EWP contribution is
the same as that of the rest of the $T+C$ amplitude, permitting the
calculation of the EWP correction.

\begin{figure}
\begin{center}
\includegraphics[height=3.5in]{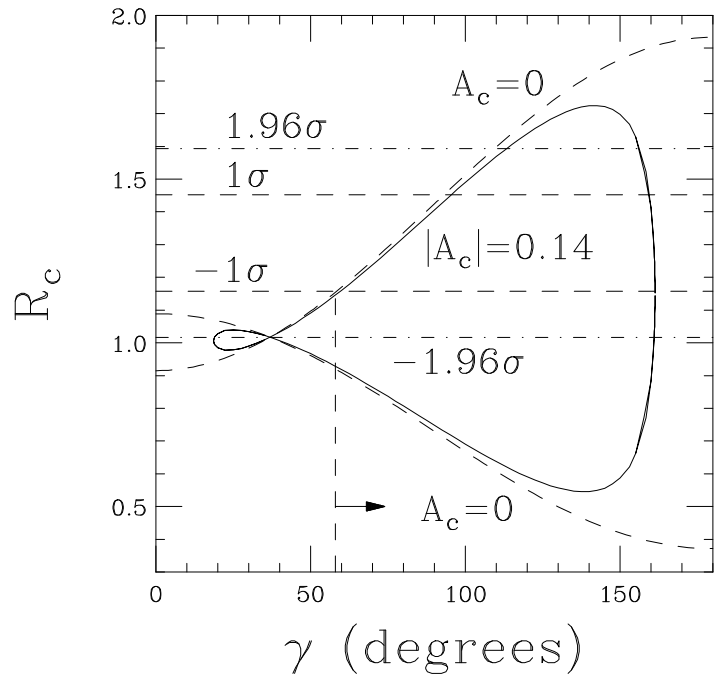}
\caption{Behavior of $R_c$ for $r_c = 0.217$ ($1 \sigma$ upper limit) and
$A_c = 0$ (dashed curves) or $|A_c| = 0.14$ (solid curve) as a function of the
weak phase $\gamma$. Horizontal dashed lines denote $\pm 1 \sigma$ experimental
limits on $R_c$, while dotdashed lines denote 95\% c.l. ($ \pm 1.96 \sigma$)
limits.  Upper branches of curves correspond to $\cos \delta_c(\cos \gamma -
\delta_{EW}) < 0$, where $\delta_c$ is a strong phase, while lower branches
correspond to $\cos \delta_c(\cos \gamma - \delta_{EW}) > 0$.  Here we have
taken $\delta_{EW} = 0.80$ (its $1 \sigma$ upper limit), which
leads to the most conservative bound on $\gamma$.
\label{fig:Rc}}
\end{center}
\end{figure}

New data on branching ratios and CP asymmetries permit an update of previous
analyses.\cite{GR02,NR}  One makes use of the ratios
\beq
R_c \equiv \frac{2 \gb(B^+ \to K^+ \pi^0)}{\gb(B^+ \to K^0 \pi^+)} = 1.30
\pm 0.15~~,~~~
r_c = |(T+C)/P| = 0.20 \pm 0.02~~,
\eeq
and
\beq
A_c \equiv \frac{\Gamma(B^- \to K^- \pi^0) - \Gamma(B^+ \to K^+ \pi^0)}
{\gb(B^+ \to K^0 \pi^+)} = 0.05 \pm 0.09~~~.
\eeq
One must also use an estimate \cite{NR} of the electroweak penguin parameter
$\delta_{\rm EW} = 0.65 \pm 0.15$.  One obtains the most conservative (i.e.,
weakest) bound on $\gamma$ for the maximum values of $r_c$ and $\delta_{\rm
EW}$.\cite{GR02}  The resulting plot is shown in Fig.\ \ref{fig:Rc}.  One
obtains a bound at the $1 \sigma$ level very similar to that in the previous
case:  $\gamma \gs 58^\circ$.

\section{Summary}

The process $B^0 \to J/\psi K_S$ has provided spectacular confirmation of the
Kobayashi-Maskawa theory of CP violation, measuring $\phi_1 = \beta$ to a few
degrees.  Now one is entering the territory of more difficult measurements.

The decay $B^0 \to \pi^+ \pi^-$ has great potential for giving useful
information on $\phi_2 = \alpha$.  One needs either a measurement of
${\cal B}(B^0 \to \pi^0 \pi^0)$,\cite{GL} probably at the $10^{-6}$ level
(present limits \cite{Bebrs,Babrs,CLbrs} are several times that), or a
better estimate of the tree amplitude from $B \to \pi l \nu$.\cite{LR}  As
for the BaBar and Belle experimental CP asymmetries,\cite{Bapipi,Bepipi}
they will eventually converge to one another, as did the initial measurements
of $\sin 2 \phi_1$ using $B^0 \to J/\psi K_S$.

The $B \to \phi K_S$ decay can display new physics via special $\bar b \to \bar
s s \bar s$ operators or effects on the $\bar b \to \bar s$ penguin.  Some
features of any new amplitude can be extracted from the data in a
model-independent way if one uses both rate and asymmetry
information.\cite{CR03}

The rate for $B \to \eta' K_S$ is not a problem for the standard model if one
allows for a modest flavor-singlet penguin contribution in addition to the
standard penguin amplitude.  The CP asymmetries for this process are in accord
with the expectations of the standard model at the $1 \sigma$ level or
better.  Effects of the singlet penguin amplitude may also be visible
elsewhere, for example in $B^+ \to p \bar p K^+$.

Various ratios of $B \to K \pi$ rates, when combined with information on
CP asymmetries, show promise for constraining phases in the CKM matrix.
These tests have shown a steady improvement in accuracy since the asymmetric
$B$ factories have been operating, and one expects further progress as the
instantaneous and accumulated luminosities increase.  In the longer term,
hadron colliders may provide important contributions.
 
\section*{Acknowledgments}

I thank C.-W. Chiang, M. Gronau, and Z. Luo for
collaborations on the subjects mentioned here; Tom Browder,
Steve Olsen, Sandip Pakvasa, Xerxes Tata, and San Fu Tuan for discussions, and
the Physics Department of the University of Hawaii for hospitality.
This work was supported in part by the United
States Department of Energy through Grant No.\ DE FG02 90ER40560.

\section*{References}

\def \ajp#1#2#3{Am.\ J. Phys.\ {\bf#1}, #2 (#3)}
\def \apny#1#2#3{Ann.\ Phys.\ (N.Y.) {\bf#1}, #2 (#3)}
\def \app#1#2#3{Acta Phys.\ Polonica {\bf#1}, #2 (#3)}
\def \arnps#1#2#3{Ann.\ Rev.\ Nucl.\ Part.\ Sci.\ {\bf#1}, #2 (#3)}
\def \art{and references therein}
\def \cmts#1#2#3{Comments on Nucl.\ Part.\ Phys.\ {\bf#1}, #2 (#3)}
\def \cn{Collaboration}
\def \cp89{{\it CP Violation,} edited by C. Jarlskog (World Scientific,
Singapore, 1989)}
\def \efi{Enrico Fermi Institute Report No.\ }
\def \epjc#1#2#3{Eur.\ Phys.\ J. C {\bf#1}, #2 (#3)}
\def \f79{{\it Proceedings of the 1979 International Symposium on Lepton and
Photon Interactions at High Energies,} Fermilab, August 23-29, 1979, ed. by
T. B. W. Kirk and H. D. I. Abarbanel (Fermi National Accelerator Laboratory,
Batavia, IL, 1979}
\def \hb87{{\it Proceeding of the 1987 International Symposium on Lepton and
Photon Interactions at High Energies,} Hamburg, 1987, ed. by W. Bartel
and R. R\"uckl (Nucl.\ Phys.\ B, Proc.\ Suppl., vol.\ 3) (North-Holland,
Amsterdam, 1988)}
\def \ib{{\it ibid.}~}
\def \ibj#1#2#3{~{\bf#1}, #2 (#3)}
\def \ichep72{{\it Proceedings of the XVI International Conference on High
Energy Physics}, Chicago and Batavia, Illinois, Sept. 6 -- 13, 1972,
edited by J. D. Jackson, A. Roberts, and R. Donaldson (Fermilab, Batavia,
IL, 1972)}
\def \ijmpa#1#2#3{Int.\ J.\ Mod.\ Phys.\ A {\bf#1}, #2 (#3)}
\def \ite{{\it et al.}}
\def \jhep#1#2#3{JHEP {\bf#1}, #2 (#3)}
\def \jpb#1#2#3{J.\ Phys.\ B {\bf#1}, #2 (#3)}
\def \lg{{\it Proceedings of the XIXth International Symposium on
Lepton and Photon Interactions,} Stanford, California, August 9--14 1999,
edited by J. Jaros and M. Peskin (World Scientific, Singapore, 2000)}
\def \lkl87{{\it Selected Topics in Electroweak Interactions} (Proceedings of
the Second Lake Louise Institute on New Frontiers in Particle Physics, 15 --
21 February, 1987), edited by J. M. Cameron \ite~(World Scientific, Singapore,
1987)}
\def \kdvs#1#2#3{{Kong.\ Danske Vid.\ Selsk., Matt-fys.\ Medd.} {\bf #1},
No.\ #2 (#3)}
\def \ky85{{\it Proceedings of the International Symposium on Lepton and
Photon Interactions at High Energy,} Kyoto, Aug.~19-24, 1985, edited by M.
Konuma and K. Takahashi (Kyoto Univ., Kyoto, 1985)}
\def \mpla#1#2#3{Mod.\ Phys.\ Lett.\ A {\bf#1}, #2 (#3)}
\def \nat#1#2#3{Nature {\bf#1}, #2 (#3)}
\def \nc#1#2#3{Nuovo Cim.\ {\bf#1}, #2 (#3)}
\def \nima#1#2#3{Nucl.\ Instr.\ Meth. A {\bf#1}, #2 (#3)}
\def \np#1#2#3{Nucl.\ Phys.\ {\bf#1}, #2 (#3)}
\def \npbps#1#2#3{Nucl.\ Phys.\ B Proc.\ Suppl.\ {\bf#1}, #2 (#3)}
\def \os{XXX International Conference on High Energy Physics, Osaka, Japan,
July 27 -- August 2, 2000}
\def \PDG{Particle Data Group, K. Hagiwara \ite, \prd{66}{010001}{2002}}
\def \pisma#1#2#3#4{Pis'ma Zh.\ Eksp.\ Teor.\ Fiz.\ {\bf#1}, #2 (#3) [JETP
Lett.\ {\bf#1}, #4 (#3)]}
\def \pl#1#2#3{Phys.\ Lett.\ {\bf#1}, #2 (#3)}
\def \pla#1#2#3{Phys.\ Lett.\ A {\bf#1}, #2 (#3)}
\def \plb#1#2#3{Phys.\ Lett.\ B {\bf#1}, #2 (#3)}
\def \pr#1#2#3{Phys.\ Rev.\ {\bf#1}, #2 (#3)}
\def \prc#1#2#3{Phys.\ Rev.\ C {\bf#1}, #2 (#3)}
\def \prd#1#2#3{Phys.\ Rev.\ D {\bf#1}, #2 (#3)}
\def \prl#1#2#3{Phys.\ Rev.\ Lett.\ {\bf#1}, #2 (#3)}
\def \prp#1#2#3{Phys.\ Rep.\ {\bf#1}, #2 (#3)}
\def \ptp#1#2#3{Prog.\ Theor.\ Phys.\ {\bf#1}, #2 (#3)}
\def \rmp#1#2#3{Rev.\ Mod.\ Phys.\ {\bf#1}, #2 (#3)}
\def \rp#1{~~~~~\ldots\ldots{\rm rp~}{#1}~~~~~}
\def \rpp#1#2#3{Rep.\ Prog.\ Phys.\ {\bf#1}, #2 (#3)}
\def \sing{{\it Proceedings of the 25th International Conference on High Energy
Physics, Singapore, Aug. 2--8, 1990}, edited by. K. K. Phua and Y. Yamaguchi
(Southeast Asia Physics Association, 1991)}
\def \slc87{{\it Proceedings of the Salt Lake City Meeting} (Division of
Particles and Fields, American Physical Society, Salt Lake City, Utah, 1987),
ed. by C. DeTar and J. S. Ball (World Scientific, Singapore, 1987)}
\def \slac89{{\it Proceedings of the XIVth International Symposium on
Lepton and Photon Interactions,} Stanford, California, 1989, edited by M.
Riordan (World Scientific, Singapore, 1990)}
\def \smass82{{\it Proceedings of the 1982 DPF Summer Study on Elementary
Particle Physics and Future Facilities}, Snowmass, Colorado, edited by R.
Donaldson, R. Gustafson, and F. Paige (World Scientific, Singapore, 1982)}
\def \smass90{{\it Research Directions for the Decade} (Proceedings of the
1990 Summer Study on High Energy Physics, June 25--July 13, Snowmass, Colorado),
edited by E. L. Berger (World Scientific, Singapore, 1992)}
\def \tasi{{\it Testing the Standard Model} (Proceedings of the 1990
Theoretical Advanced Study Institute in Elementary Particle Physics, Boulder,
Colorado, 3--27 June, 1990), edited by M. Cveti\v{c} and P. Langacker
(World Scientific, Singapore, 1991)}
\def \yaf#1#2#3#4{Yad.\ Fiz.\ {\bf#1}, #2 (#3) [Sov.\ J.\ Nucl.\ Phys.\
{\bf #1}, #4 (#3)]}
\def \zhetf#1#2#3#4#5#6{Zh.\ Eksp.\ Teor.\ Fiz.\ {\bf #1}, #2 (#3) [Sov.\
Phys.\ - JETP {\bf #4}, #5 (#6)]}
\def \zpc#1#2#3{Zeit.\ Phys.\ C {\bf#1}, #2 (#3)}
\def \zpd#1#2#3{Zeit.\ Phys.\ D {\bf#1}, #2 (#3)}

\end{document}